\documentclass[11pt,aps,prb,unsortedaddress]{revtex4}   
\usepackage{graphicx}

\usepackage{amsmath}    % need for subequations
\usepackage{graphicx}   % for figures
\usepackage{placeins}
\usepackage{epsfig}
\usepackage{hyperref}
\usepackage{subfigure}

\newcommand{\ajp}{AJP}  % example of a definition of a macro

\newcommand{\apjl}{Astrophys.~J.~L.}
\newcommand{\apjs}{Astrophys.~J.~S.}
\newcommand{\mnras}{Mon.~Not.~Royal~Ast.~Soc.}

\def\be{\begin{equation}}
\def\ee{\end{equation}}
\def\bi{\begin{itemize}}
\def\ei{\end{itemize}}
\def\ben{\begin{enumerate}}
\def\een{\end{enumerate}}
\def\i{\item{}}

\begin{document}

\title{An acoustical analogue of a galactic-scale gravitational-wave detector}

\author{Michael T. Lam}
\email{michael.lam@mail.wvu.edu}   %optional
\affiliation{Department of Physics and Astronomy, West Virginia University, P.O. Box 6315, Morgantown, WV 26506, USA}
\affiliation{Center for Gravitational Waves \& Cosmology,
West Virginia University, Chestnut Ridge Research Building, 
Morgantown, WV 26505, USA}

\author{Joseph D.~Romano}
\email{joseph.romano@utrgv.edu}   %optional
\affiliation{Department of Physics and Astronomy,
University of Texas Rio Grande Valley,
Brownsville, TX 78520, USA}

\author{Joey S.~Key}
\email{joeykey@uw.edu}   %optional
\affiliation{Division of Physical Sciences,
University of Washington Bothell,
Bothell, WA 98011, USA}

\author{Marc Normandin}
\email{marc.normandin@utrgv.edu}   %optional
\affiliation{Department of Physics and Astronomy,
University of Texas Rio Grande Valley,
Brownsville, TX 78520, USA}

\author{Jeffrey S.~Hazboun}
\email{jeffrey.hazboun@utrgv.edu}   %optional
\affiliation{Center for Advanced Radio Astronomy,
University of Texas Rio Grande Valley,
Brownsville, TX 78520, USA}

\date{\today}

\begin{abstract}
By precisely monitoring the ``ticks" of Nature's most
precise clocks (millisecond pulsars), scientists are trying to 
detect the ``ripples in spacetime" (gravitational waves)
produced by the inspirals of supermassive black holes
in the centers of distant merging galaxies.
Here we describe a relatively simple
demonstration that uses two metronomes 
and a microphone to illustrate several techniques used 
by pulsar astronomers to 
search for gravitational waves.
An adapted version of this demonstration could be used 
as an instructional laboratory investigation at the 
undergraduate level.
\end{abstract}

\maketitle

\section{Introduction}
\label{s:introduction}

A {\em pulsar timing array} is a galactic-scale 
gravitational-wave detector, 
which can be used to search for gravitational 
waves from the inspiral of supermassive black-hole binaries 
(of order $10^9$~solar masses) in the centers of distant 
galaxies\citep{PTACQG,Detweiler1979,hd1983}.
The array consists of a set of galactic millisecond
{\em pulsars}---rapidly-rotating neutron stars, which have 
masses of order the mass of the Sun and magnetic fields 
of order a billion times stronger than that of the Earth\citep{handbook}.
Millisecond pulsars rotate nearly a thousand times
each second (faster than a kitchen blender), 
emitting a narrow beam of radio waves 
along the magnetic axes that sweep across the sky similar to a revolving beacon on top of a lighthouse.
If this radio beam crosses our line of sight to the
pulsar, a radio telescope on Earth will observe pulses
of radiation, which arrive with a regularity that rivals 
(or even exceeds) that of the best atomic clocks\citep{Hobbs+2012}.

By precisely monitoring the pulse arrival times,
radio astronomers can determine what the rotation 
period of the pulsar is, how the rotation is
slowing down, whether the pulsar is orbiting 
a companion star, as well as how the
interstellar medium affects the propagation of the pulses\citep{handbook}.
The difference between the {\em measured} times of 
arrival and the {\em expected} times 
of arrival (taking all of these effects into account) 
are called {\em timing residuals}.
If the pulsar timing model is good, the residuals 
should be randomly scattered around zero with a
root-mean-square (rms) amplitude determined by 
measurement noise in the radio receiver and 
statistical fluctuations in the pulses themselves.
The residuals for an individual pulsar may be 
correlated in time\citep{cd1985,IPTADR1noise,NG9EN,CordesCQG} (so-called red noise), but 
residuals associated with different Earth-pulsar 
baselines should not be correlated with one another
in the absence of any common external influence.
Deviations from this expected behavior could be 
due to either an incomplete timing model 
(e.g., not realizing that the pulsar is in a 
binary) or the presence of gravitational waves\citep{ERA}.

\begin{figure}[hbtp!]
\begin{center}
\includegraphics[clip=true, angle=0, width=0.6\textwidth]{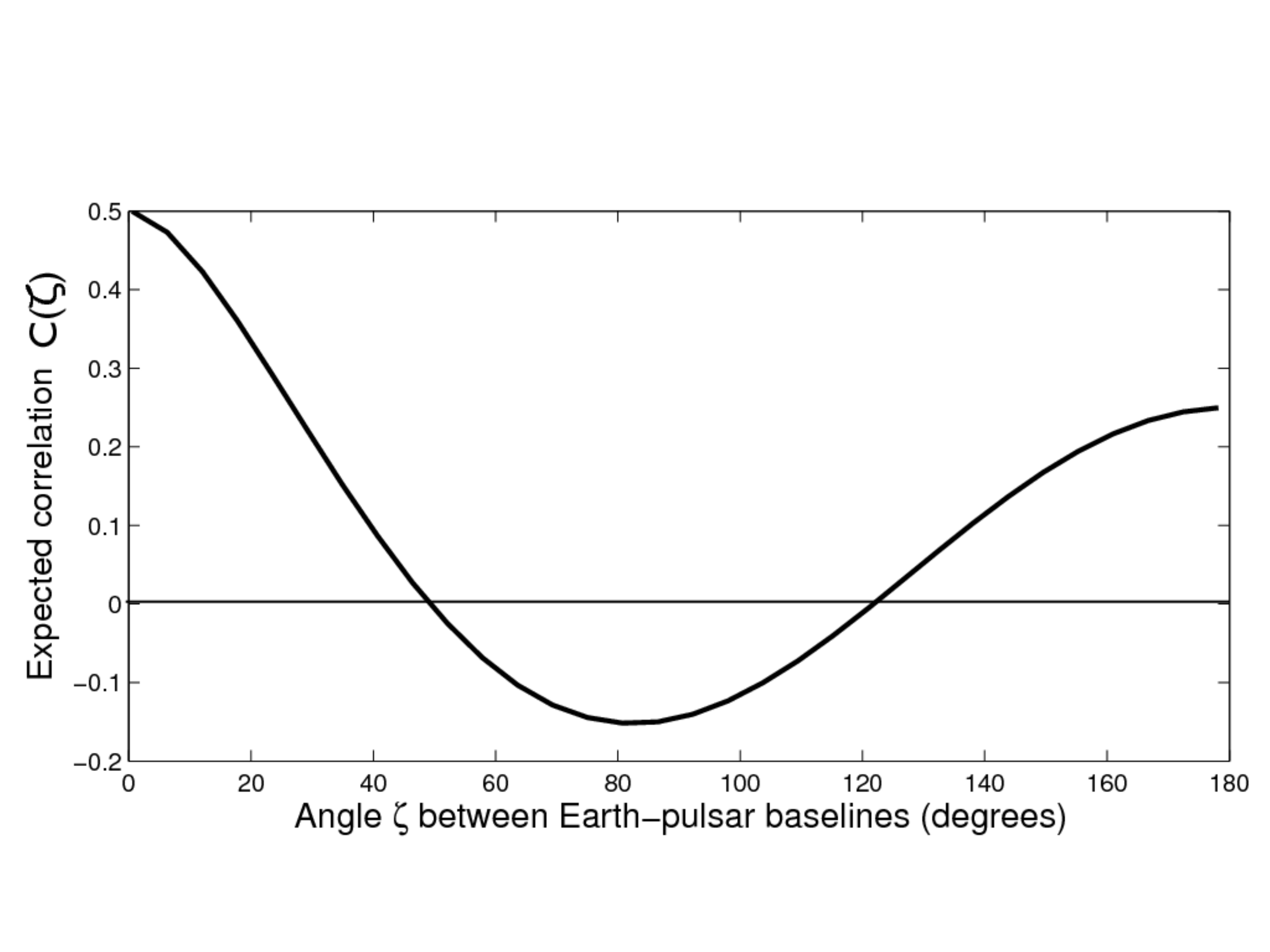}
\caption{Expected correlation coefficient between the 
timing residuals for a pair of Earth-pulsar baselines
separated by angle $\zeta$.}
\label{f:HDcurve}
\end{center}
\end{figure}

A gravitational wave passing between the Earth 
and a pulsar will stretch and squeeze space 
transverse to its motion, slightly advancing
or retarding the arrival times of the individual
pulses\citep{ew1975}. 
Unlike the measurement noise or intrinsic 
pulsar timing noise discussed above, the modulation 
of pulse arrival times induced by a gravitational
wave will be {\em correlated} across different 
pulsars in the array, due to its common influence 
in the vicinity of the Earth. 
Moreover, this correlation will have a very 
specific dependence on the angle between a pair of 
Earth-pulsar baselines, the so-called 
{\em Hellings and Downs curve}\citep{hd1983} shown in 
Figure~\ref{f:HDcurve}.
The detection of this expected correlation in the timing residuals 
from an array of pulsars would be evidence for the presence of
gravitational waves, similar to the recent detections by the 
advanced LIGO and Virgo detectors\cite{GW150914,LIGOPapers,GW170817}.

\subsection{Metronomes and microphones}
\label{s:metronomes_microphones}

In order to illustrate how gravitational-wave astronomers 
are using correlation methods to search for gravitational waves, 
we have developed a demonstration using metronomes and 
a microphone, which serves as an {\em acoustical analogue} 
of a pulsar timing array.
In this demonstration, radio pulses from an array of 
Galactic pulsars are represented by ticks of an array of metronomes
(only two metronomes are needed for this demonstration);
radio receivers on Earth are represented by a single microphone;
and the passage of a gravitational wave is represented by the 
motion of the microphone around its nominal position.
The analogy is not perfect as the motion of the microphone
does not represent a wave of any kind, and the correlations 
that it induces have a different angular dependence than 
that induced by a real gravitational wave\citep{jr2015}.
But what is important is that there {\em are} correlations,
as the microphone motion modulates the arrival 
times of the metronome pulses by changing the distance 
between the metronomes and the microphone.
And although the angular dependence of the correlations 
for the microphone motion is different than that for 
gravitational waves, it is, nonetheless, a specific 
function of the angle between a pair of microphone-metronome 
baselines, which can be calculated theoretically and
also verified experimentally by doing the demonstration.

In the following sections, we will describe the 
metronome-microphone demonstration in detail.
In Section~\ref{s:hardware_software}, 
we describe the specific hardware (i.e.,
metronomes and microphone) and software routines that 
we use to do the analysis.
In Section~\ref{s:techniques}, 
we list the techniques used in real pulsar
timing analyses that are illustrated by the demonstration.
They can be thought of as the \emph{learning outcomes} for the 
demonstration.
In Sections~\ref{s:analysis1} and \ref{s:analysis2}, 
we discuss the two main parts of the demonstration 
(the single-metronome and double-metronome analyses), 
listing the steps needed to perform the analysis and 
the function of the graphical user interface (GUI) buttons used to execute each step.
In Section~\ref{s:discussion}, 
we conclude with a discussion of some caveats
and possible improvements to the demonstration, and 
how it might be adapted for use in the 
collection of high school and undergraduate 
laboratory\citep{Rubbo+2007,Newburgh2008,fsa2015,Burko2017} and classroom\citep{Farr+2012,Kassner2015,Mathur+2017,Kaur+2017,Hilborn2018} investigations centered 
around understanding gravitational physics. %FOO
[Sample data files and analysis routines are available
for download from URL \url{http://github.com/josephromano/pta-demo}.]

%%%%%%%%%%%%%%%%%%%%%%%%%%%%%%%%%%%%%%%%%%%%%%%%%%%%%%%
\section{Required hardware and software}
\label{s:hardware_software}

\begin{figure}[hbtp!]
\begin{center}
\includegraphics[clip=true, angle=0, width=.25\textwidth]{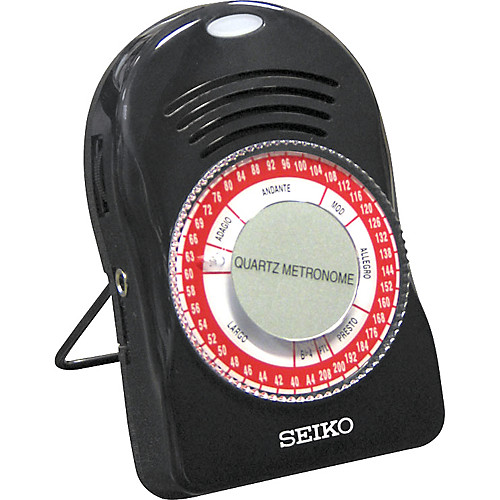}
\includegraphics[clip=true, angle=0, width=.25\textwidth]{metronome}
\includegraphics[clip=true, angle=0, width=.3\textwidth]{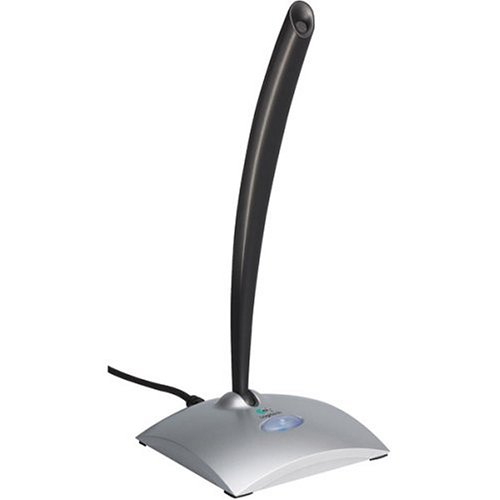}
\caption{Two Seiko metronomes and one Logitech USB noise-canceling
microphone used for the demonstration.}
\label{f:metronome-microphone}
\end{center}
\end{figure}

The metronome-microphone pulsar-timing-array 
demonstration requires two metronomes.
Our preferred choice is Seiko model SQ50-V 
quartz metronomes 
(Figure~\ref{f:metronome-microphone}), 
as this model has adjustable 
beats-per-minute (bpm) up to 208 bpm, 
adjustable volume, and 
two different tempo sounds---mode $a$ and mode $b$,
with mode $b$ having a slightly higher pitch.
Having two modes is helpful in distinguishing the
pulses from the individual metronomes when both 
metronomes are on simultaneously, since the pulse 
shapes (profiles) are different.

One also needs some type of microphone, either an 
external USB microphone or an internal microphone,
connected to a laptop that is set up to run the 
relevant data analysis routines (described below).
We have found that the internal microphone on 
a MacBook Pro works best since it has 
ambient noise reduction, although it is somewhat 
inconvenient to physically move the laptop to 
simulate the passage of a gravitational wave.
(We move the microphone is a small circle of 
radius $\sim\!10~{\rm cm}$ at constant speed, 
for reasons we will describe below.)
We have also used a Logitech USB Desktop 
noise-canceling microphone (Figure~\ref{f:metronome-microphone}), 
which is a little easier to maneuver.

\begin{figure}[hbtp!]
\begin{center}
\includegraphics[clip=true, angle=90, width=\textwidth]{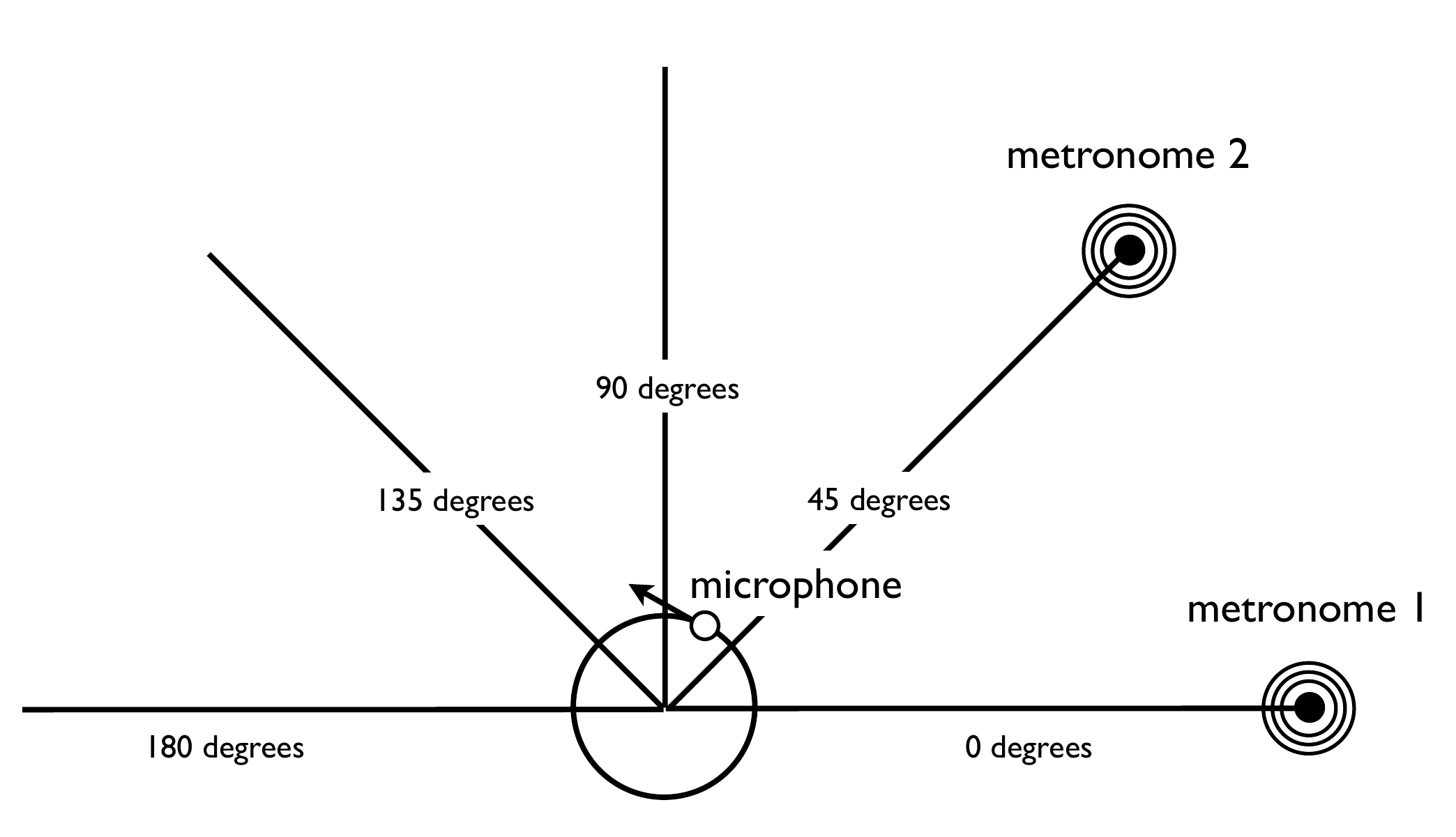}
\caption{Schematic diagram showing the location of the microphone 
and metronomes for the different analyses.
The stationary microphone is located at the origin;
the moving microphone undergoes uniform circular motion, 
indicated by the counter-clockwise circle.
Metronome 2 is placed at angular location $45^\circ$ in this figure,
but will be placed at the other angular locations for different parts
of the demonstration.}
\label{f:setup}
\end{center}
\end{figure}

\begin{figure}[hbtp!]
\begin{center}
\includegraphics[clip=true, angle=90, width=\textwidth]{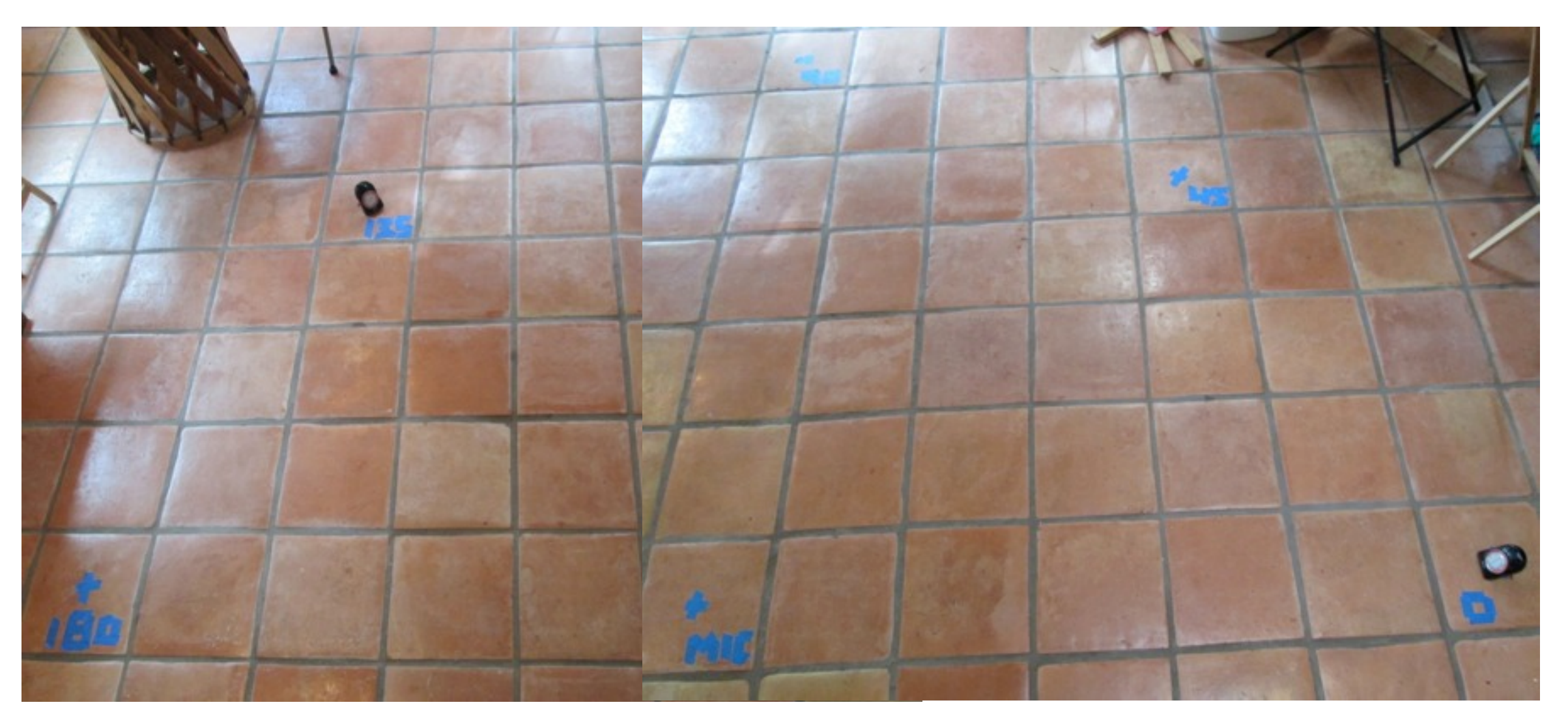}
\caption{Photograph of an actual setup for data taking.
Metronome 1 is shown at angular location $0^\circ$; 
metronome 2 at angular location $135^\circ$.
The separation between the microphone (located at the origin) and the
metronomes at the edge of a semi-circle is approximately 5~feet.}
\label{f:actualsetup}
\end{center}
\end{figure}

In addition, one needs an open space covering an area of 
about $10~{\rm ft}\times 5~{\rm ft}$ 
for the placement of the two metronomes and microphone.
A schematic diagram of the setup is shown in
Figure~\ref{f:setup}.
A photograph of an actual real-world setup used to take the
data is shown in Figure~\ref{f:actualsetup}.

We have written Python-based routines to do the 
relevant data analysis calculations.
These includes routines for:
(i) audio recording and play back of metronome pulses, 
(ii) pulse data folding, 
(iii) pulse profile calculation,
(iv) matched-filter estimation of pulse 
arrival times, 
(v) timing residual calculation, 
(vi) linear detrending of timing residuals, 
(vii) fitting of sinusoids to the timing residuals, and 
(viii) correlation coefficient calculation.
Two Python-based GUIs exist for performing the two main
data taking and data analysis tasks:  
{\tt PTAdemo1GUI.py} (for analyzing the single-metronome 
data) and
{\tt PTAdemo2GUI.py} 
(for analyzing the double-metronome data).
The two GUIs and the data analysis techniques are 
described in more detail in the following sections.

%%%%%%%%%%%%%%%%%%%%%%%%%%%%%%%%%%%%%%%%%%%%%%%%%%%%%
\section{Pulsar timing techniques illustrated by 
the demonstration}
\label{s:techniques}

The metronome-microphone demonstration is useful as 
an educational tool since it illustrates several 
techniques used in real pulsar timing analyses.
It does this in the simplified context of 
metronome pulses recorded by a microphone, which
students or the general public can more easily
identify with.
To set the stage for the analyses that will be 
described in Secs.~\ref{s:analysis1} and 
\ref{s:analysis2}, we describe below the key 
techniques illustrated by the demonstration.
%Readers not interested in the mathematical details
%can simply ignore the relevant formulae.

\subsection{Folding data}
\label{s:folding}

\begin{figure}[hbtp!]
\begin{center}
\includegraphics[clip=true, angle=0, width=\textwidth]{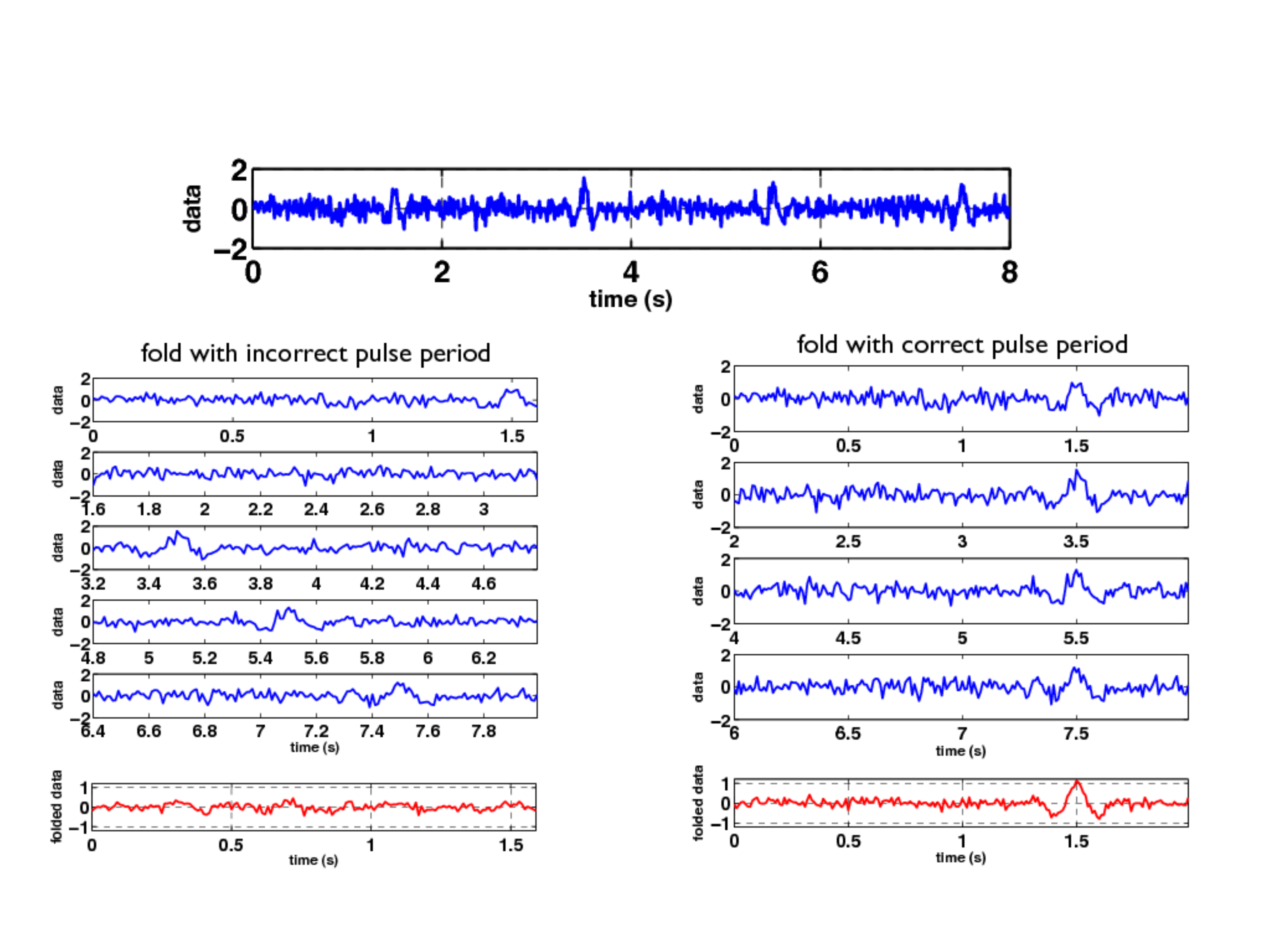}
\caption{Illustration of folding data with both an incorrect 
and the correct choice for the pulse period $T_{\rm p}$.
Top panel: Noisy time-series data with several injected pulses
having $T_{\rm p}=2~{\rm s}$.
Left column: Segments of the original time-series data each of
duration $1.6$~s (in blue), and the average of these segments
(in red).
Right column: Same as for the left column but for segments of 
duration 2~s.
Note that when the data are folded with the correct pulse period
$T_{\rm p}=2~{\rm s}$, the signal components combine coherently
and the pulse profile is easily visible in the average of the 
segments (bottom right plot).}
\label{f:folding}
\end{center}
\end{figure}

Folding data is a technique that can be used to find 
both the 
pulse period and pulse shape (profile) 
in noisy time-series data \citep{handbook}.
The time-series (of total duration $T_{\rm tot}$) 
is first split into smaller segments, each of 
duration $T$, which are then averaged together.
If $T$ matches the true pulse period $T_{\rm p}$,
then the pulse contributions in each segment 
combine coherently when the segments are summed,
while the noise contributions combine incoherently 
(positive and negative values tending to cancel one another).
If $T$ does not match the true pulse
period, the pulse contributions will effectively
cancel out when averaged against the noise.
So the basic procedure is to systematically try 
different values of $T$ until a pulse 
profile ``sticks out of the data", which will occur 
when $T$ equals $T_{\rm p}$.  
The signal-to-noise ratio of the recovered
pulse profile grows like $\sqrt{N}$, where $N$
is the number of segments or, equivalently, the 
number of individual pulses combined\citep{cs2010}.
Figure~\ref{f:folding} illustrates what happens
when you fold data with an incorrect and the 
correct pulse period.

\subsection{Matched-filter determination of pulse arrival times}
\label{s:matchedfilter}

\begin{figure}[hbtp!]
\begin{center}
\includegraphics[clip=true, angle=0, width=0.6\textwidth]{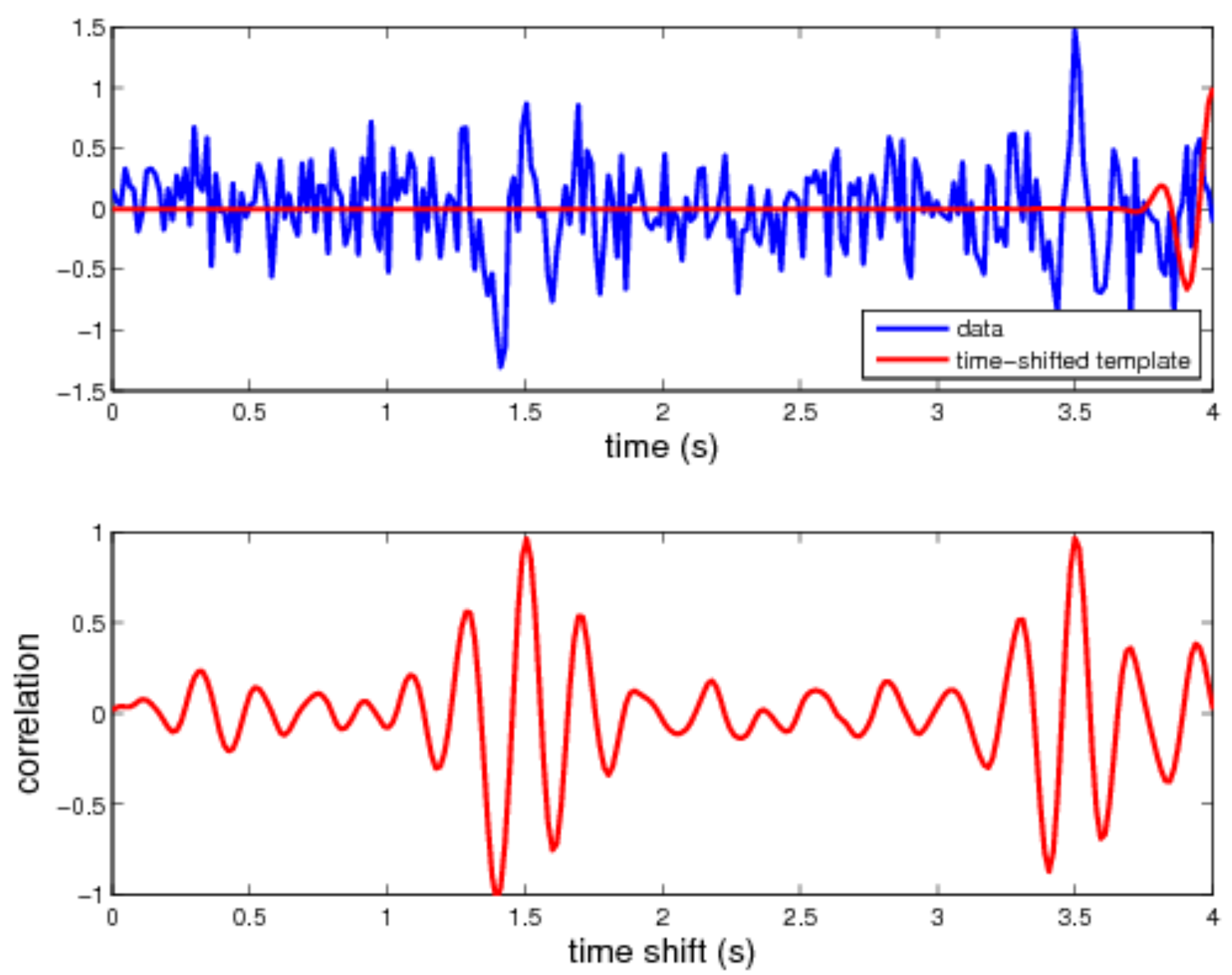}
\caption{Illustration on simulated data showing that how the
correlation function $C(\tau)$ has local maxima at the arrival 
times of the pulses.
An animation showing the calculation of $C(\tau)$ as a 
function of the timeshift of the pulse profile is available at
\url{http://github.com/josephromano/pta-demo/tree/master/code/},
with filename {\tt matchedfilterdemo.avi}.}
\label{f:matchedfilterdemo}
\end{center}
\end{figure}

The measured times of arrival (TOAs) are determined by correlating a 
time-shifted version of the pulse profile with the 
time-series data\citep{Taylor1992,NG9WN}.
Mathematically, one calculates the correlation function 
\be
C(\tau) \equiv {\cal N}\int dt\>y(t)p(t-\tau)\,,
\label{e:correlation-time}
\ee
where $y(t)$ is the timeseries and $p(t-\tau)$ is 
the pulse profile shifted forward in time by $\tau$.
(${\cal N}$ is a normalization constant, defined below.)
In the frequency domain, we have
\be
C(\tau) ={\cal N}
\int df\> 
\tilde y(f) \tilde p^*(f)e^{i2\pi f\tau}\,,
\qquad
{\cal N}
\equiv \left[\int df\> |\tilde p^*(f)|^2\right]^{-1}\,,
\label{e:correlation-freq}
\ee
where $\tilde y(f)$ and $\tilde p(f)$ are the Fourier 
transforms%
\citep{FourierFootnote}
of $y(t)$ and $p(t)$.
The correlation function $C(\tau)$ has local
maxima at the arrival times of the pulses 
(Figure~\ref{f:matchedfilterdemo}).
In what follows, 
we will denote these measured arrival times by $\tau^{\rm meas}[i]$,
where $i=1,2,\cdots$.
The normalization constant ${\cal N}$ is included  
so that the values of the correlation function at the 
measured TOAs are estimates of the amplitudes
of the pulses.

\subsection{Calculating timing residuals based on a timing model}
\label{s:timingresiduals}

Calculating the timing residuals is a simple matter of 
subtracting the expected TOAs from the measured TOAs of 
the pulses:
\be
\delta\tau[i] = \tau^{\rm meas}[i] - \tau^{\rm exp}[i]\,,
\label{e:timingresiduals}
\ee
where $i=1,2,\cdots$ labels the individual pulses.
As mentioned in Sec.~\ref{s:introduction}, for real
pulsar timing analyses the expected TOAs are determined
by a rather sophisticated timing model, which takes 
into account the rotation period of the pulsar, 
its spin-down rate, its location in the sky, etc.
But for the metronome-microphone demonstration, the 
timing model is exceedingly simple:
\be
\tau^{\rm exp}[i] = \tau^{\rm meas}[i_0] + (i-i_0)T_{\rm p}\,,
\label{e:timingmodel}
\ee
which is just the measured TOA of the pulse having the
largest correlation with the pulse profile, indexed by
$i_0$, plus integer multiples of the pulse period $T_{\rm p}$ 
of the metronome.

\subsection{Improving the timing model by removing a linear
trend in the residuals}
\label{s:detrend}

\begin{figure}[hbtp!]
\begin{center}
\subfigure[]{\includegraphics[clip=true, angle=0, width=0.49\textwidth]{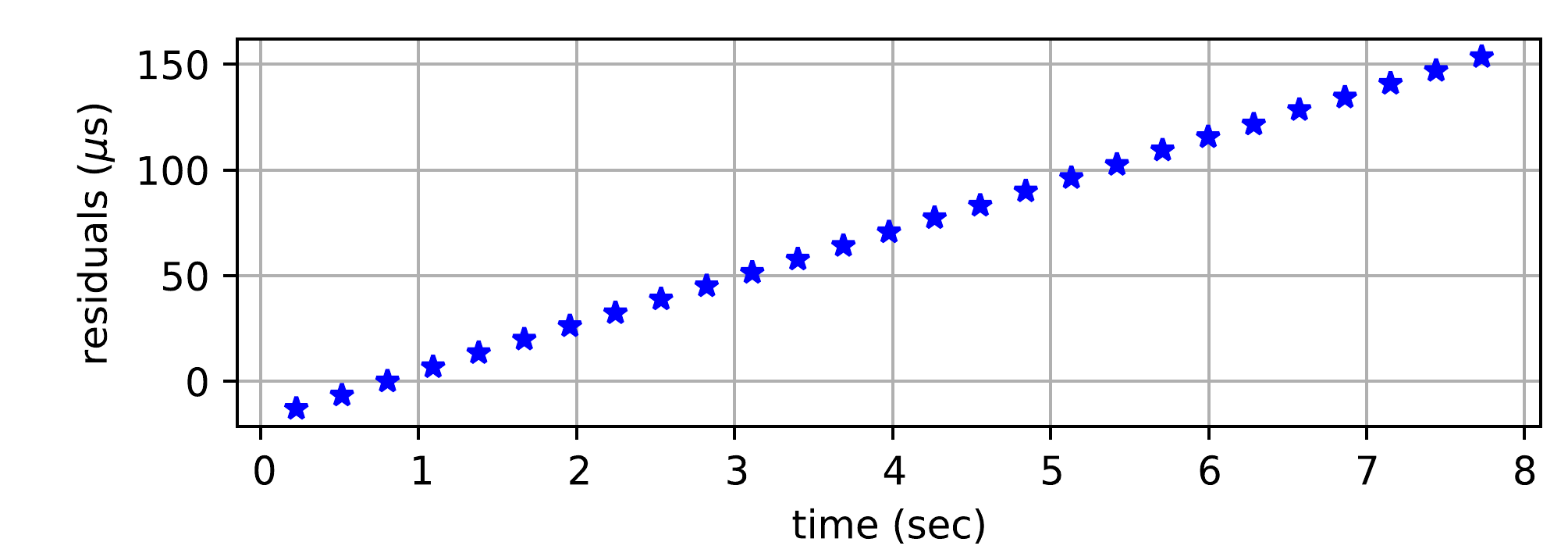}}
\subfigure[]{\includegraphics[clip=true, angle=0, width=0.49\textwidth]{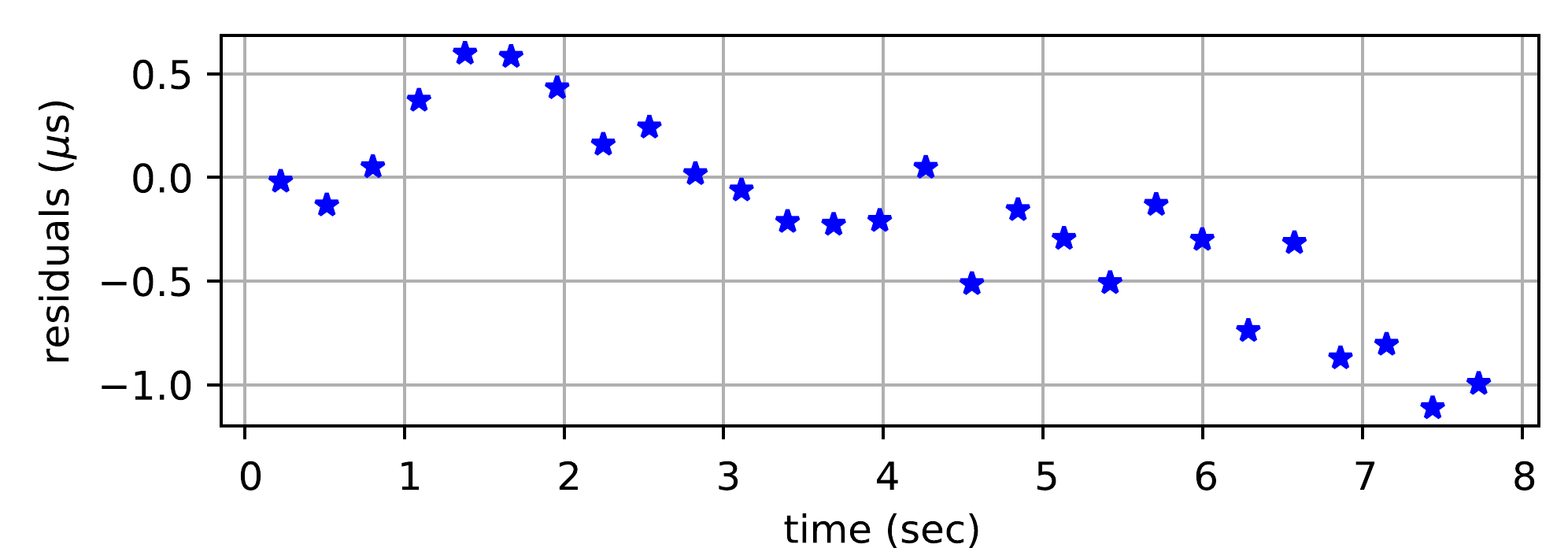}}
\caption{Residuals for metronome pulses: (a) before detrending, and (b) after detrending. Note the different scales on the $y$-axis.}
\label{f:residuals_trend_detrended}
\end{center}
\end{figure}

A linear trend in the timing residuals is an indication
that the pulse period (determined by folding the data in 
Step~1 above) is not quite right. 
This is because $\delta\tau[i]$ involves a term 
$-(i-i_0)T_{\rm p}$, and if there is an error $\epsilon$
in $T_{\rm p}$, this term grows linearly with the pulse 
number, $i$.
By fitting a line to the timing residuals and removing 
this trend, we obtain an improved estimate of the pulse
period, which we can be used for subsequent timing model 
calculations.
Figure~\ref{f:residuals_trend_detrended} show timing
residuals for metronome pulses before and after removing
a linear trend.
%
% suggest using microseconds on the y axis for both plots in this figure

\subsection{Calculating correlation coefficients between
pairs of timing residuals}
\label{s:corrcoeff}

The correlation coefficient between a pair of timing 
residuals is simply the time-averaged product of the 
timing residuals for a pair of microphone-metronome 
baselines.
More generally, we can define the {\em binned correlation function} 
\be
\begin{aligned}
&C_{12}[k]\equiv \frac{1}{N_{\rm bins}}
\sum_{i,j} \delta\tau_{1}[i]\delta\tau_{2}[j]\,,
\\
&\forall i, j
\quad{\rm s.t.}\quad
\tau^{\rm meas}_{1}[i] - \tau^{\rm meas}_{2}[j]
\in \left[k-0.5,k+0.5\right]\Delta t\,,
\end{aligned}
\ee
for two sets of timing-residuals 
$\delta\tau_1$, $\delta\tau_2$,
where $\Delta t$ is the chosen bin size.
(The condition on the indices 
$i$ and $j$ is simply that the 
difference in the measured times of arrival for 
the two timing residuals must lie in the $k$th lag bin.)
But for the purposes of the demonstration:
(i) we are interested in only the zero-lag result 
(so $k=0$), and 
(ii) to simplify the calculation, we can fit 
smooth curves $x_1(t)$, $x_2(t)$ 
to the two sets of timing residuals, for which
\be
C_{12}[0]
\quad\rightarrow\quad
\langle x_1 x_2\rangle \equiv 
\frac{1}{T_{\rm tot}}\int_0^{T_{\rm tot}} dt\> x_1(t)x_2(t)\,.
\ee
Normalizing by the rms values 
$\sqrt{\langle x_1^2\rangle}$ and $\sqrt{\langle x_2^2\rangle}$,
we get
\be
\left.
\rho_{12}\equiv
{\langle x_1 x_2\rangle}\right/\sqrt{\langle x_1^2\rangle \langle x_2^2\rangle}\,,
\label{e:corrcoeff}
\ee
for the correlation coefficient,
which takes values between $-1$ and 1.
Note that for timing residuals induced by uniform 
circular microphone motion (see Secs.~\ref{s:expectedcorrelation}
and \ref{s:UCM}), 
the best-fit smooth functions to the timing residuals will be sinusoids.

\subsection{Microphone-motion-induced timing residuals}
\label{s:microphonemotion}

Similar to calculating the response of an Earth-pulsar baseline 
to a passing gravitational wave, one can calculate the response
of a metronome timing residual to the motion of the microphone:
\be
\delta \tau_I^{\rm mic}(t)
\equiv\frac{\Delta L_I(t)}{c_{\rm s}}
=\frac{L_I(t)-L_I}{c_{\rm s}}
\simeq 
-\frac{1}{c_{\rm s}} \hat u_I\cdot \vec r(t)
\label{e:microphonemotion}
\ee
where $L_I(t)$ is the distance between 
metronome $I=1,2$ and the location of the 
microphone $\vec r(t)$ at time $t$,
$L_I$ is the nominal distance between the
metronome and the microphone pointing in direction $\hat u_I$, 
and $c_{\rm s}$ is the speed of sound in air.
This response is just the change in the 
metronome pulse propagation time due to the motion
of the microphone relative to metronome $I$.
The last (approximate) equality above is valid 
if we ignore 
correction terms of order $A/L\sim 0.1$, where 
$A\sim 10~{\rm cm}$ is the amplitude of the 
microphone motion and $L\sim 1~{\rm m}$ 
is the distance from the metronome to the 
microphone when it is at the origin.
Such an approximation amounts to ignoring the curvature
of the pulse wavefronts.

\subsection{Expected correlations in the timing
residuals induced by uniform circular motion}
\label{s:expectedcorrelation}

For the microphone undergoing uniform circular 
motion with amplitude $A$ and frequency $f_0$,
\be
\vec r(t) 
= A\left[\cos(2\pi f_0 t + \phi_0)\, \hat x + \sin(2\pi f_0 t+\phi_0)\, \hat y\right]\,,
\label{e:UCM}
\ee
it follows immediately that
\be
\delta \tau_I^{\rm mic}(t) 
\simeq -\frac{A}{c_{\rm s}}\cos(2\pi f_0 t+\phi_0-\theta_I)\,,
\label{e:microphonemotion-approx}
\ee
where $\theta_I$ is the angle that the location of metronome 
$I$ makes with respect to the $x$-axis, and where (as before) 
we have ignored the higher-order correction terms to the residual.
Using the trigonometric identity 
\be
\cos A\cos B =\frac{1}{2}\left[
\cos(A+B) + \cos(A-B)
\right]\,,
\ee
it is fairly easy to show that the time-averaged 
correlation coefficient of
the microphone-induced timing residuals for metronomes 1 and 2 is
\be
\rho_{12}\simeq \cos\zeta\,,
\label{e:rho12}
\ee
where $\zeta\equiv \theta_1-\theta_2$ is the angle between 
the two microphone-metronome baselines.
This equality is correct to order $(A/L)^2$.
This dependence of the correlation coefficient on the angle
between the two metronomes is what we confirm experimentally
with the double-metronome analysis described in 
Sec.~\ref{s:analysis2}.
Since the timing residuals 
for the two metronomes typically 
are evaluated at different times, we actually correlate the 
best-fit sinusoids to the timing residuals, as described
in Sec.~\ref{s:corrcoeff}.

\subsection{Justification of the choice of uniform circular motion
for the microphone}
\label{s:UCM}

In principle, we could move the microphone in any way 
whatsoever, and we would still see correlations in the 
timing residuals associated with the two metronomes.
But the form of the expected correlation will be more
complicated than the simple $\rho_{12} \simeq\cos\zeta$
dependence that we found above.
For example, if instead of uniform circular motion we let
the microphone swing back-and-forth sinusoidally 
in a plane (i.e., along some line in the $xy$-plane), 
then the correlation coefficient will also depend on 
the angle that this plane makes with the $x$-axis.
Said another way, uniform circular motion has the nice 
property that the $x$ and $y$ components of its motion 
are statistically equivalent, but uncorrelated with one 
another.
It turns out that this is also the assumption that 
goes into the calculation of the Hellings and Downs 
correlation curve (Figure~\ref{f:HDcurve}) for the 
real pulsar-timing gravitational-wave case---i.e.,
the Hellings and Downs curve is derived under the
assumption that the pulsar timing array is responding 
to a stochastic background of gravitational waves 
that is both {\em isotropic} (no preferred direction)
and {\em unpolarized} (statistically equivalent
and uncorrelated linear polarization components)\citep{hd1983}.

From a different perspective, 
the effect of uniform circular motion 
on the timing residuals is exactly what one sees in real 
pulsar-timing timing residuals if the yearly orbital 
motion of the Earth around the Sun is not taken into 
account in the timing models for the pulsars.

%%%%%%%%%%%%%%%%%%%%%%%%%%%%%%%%%%%%%%%%%%%%%%%%%%%
\section{Single-metronome analysis}
\label{s:analysis1}

The purpose of the single-metronome analysis is to
calculate the pulse period and pulse profile for 
each metronome separately in the absence of microphone motion.
The pulse periods and pulse profiles calculated here
can be considered as 
{\em reference} periods and profiles, to be used 
as inputs for the double-metronome analysis, where 
both metronomes are running simultaneously and the 
microphone is moving when it is recording pulses.

A screenshot of the GUI for this analysis, 
{\tt PTAdemo1GUI.py}, is shown in Figure~\ref{f:PTAdemo1GUI}.
The GUI has space for plots of:
(i) pulses from the individual metronomes,
(ii) pulse profiles (obtained by folding the pulse data),
and 
(iii) timing residuals.
The GUI also has several text entry fields and buttons, 
whose functions are described below:
\begin{figure}[hbtp!]
\begin{center}
\includegraphics[clip=true, angle=0, width=\textwidth]{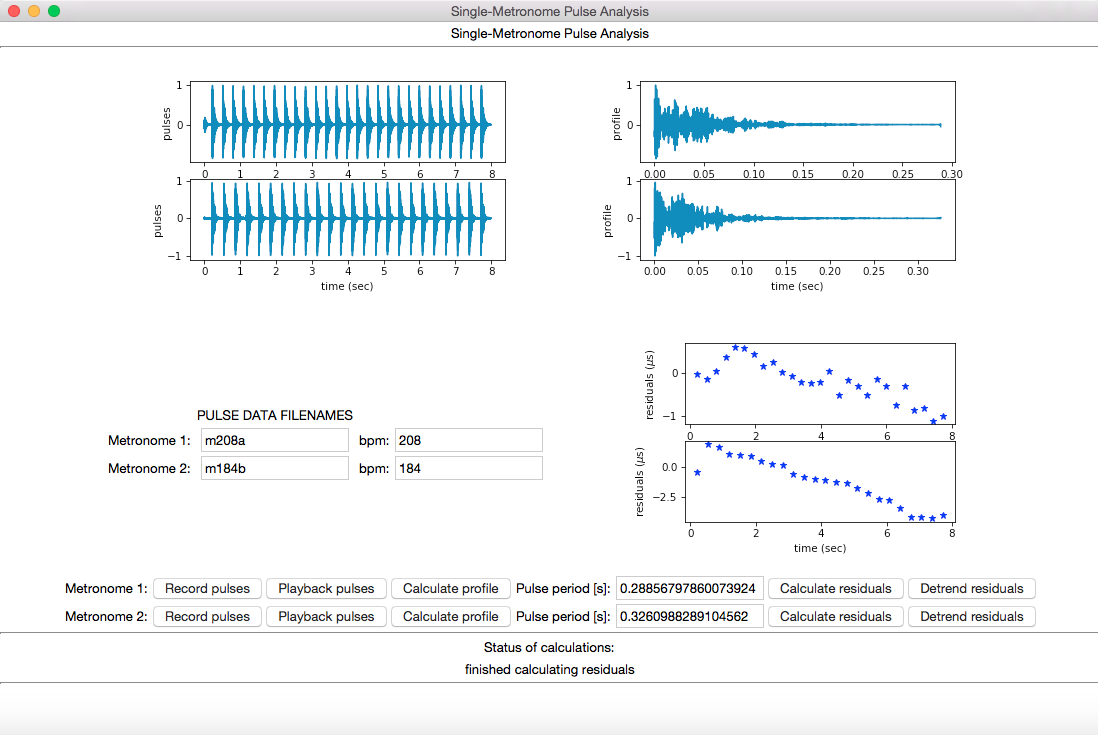}
\caption{Python GUI for the single-metronome analysis.
Plots, text entry fields, and buttons are described in the main text.}
\label{f:PTAdemo1GUI}
\end{center}
\end{figure}

\begin{description}

\i {\bf Record pulses}:
Records audio pulse data from a metronome, and save the corresponding
timeseries to an ascii {\tt .txt} file with file prefix specified by 
the {\tt PULSE DATA FILENAMES} text entry boxes (default {\tt m208a} or {\tt m184b}).
The {\tt bpm} text entry boxes have the beats-per-minute settings for 
the two metronomes. 
The pulse recording routine is hard-coded to record 8~seconds of data.

\i {\bf Playback pulses}:
Plays back and plots the audio pulse data saved in the ascii files, 
again defaulted to {\tt m208a.txt} or {\tt m184b.txt}.

\i {\bf Calculate profile}:
Either 
(i) calculates the pulse period $T_{\rm p}$ and pulse profile directly 
by folding the metronome data,
and then saves the profile to the file 
{\tt m208a\_profile.txt} or {\tt m184b\_profile.txt}, or 
(ii) reads in the pulse profile data that has already been saved 
in these files.
Method (i) is used only the first time the analysis is run.
If the pulse profiles are read-in from the files 
{\tt m208a\_profile.txt} and {\tt m184b\_profile.txt}, the text entry 
boxes for the pulse periods need to be entered by hand.
For both cases, the pulse profile is plotted from 0 to $T_{\rm p}$.

\i {\bf Calculate residuals}:
Calculates the timing residuals by subtracting the expected 
TOAs from the measured TOAs of the pulses as 
described in 
Secs.~\ref{s:timingresiduals} and \ref{s:matchedfilter}.

\i {\bf Detrend residuals}:
Improves the estimate of the pulse period for 
a metronome by removing a linear trend 
from the timing residuals as described in Sec.~\ref{s:detrend}.
Detrending may change the estimate of the pulse period
by 1-2 microseconds.
The updated period is displayed in the {\tt Pulse period}
text entry box.

\end{description}

\subsection{Steps for doing the analysis}

\ben

\i Record pulses from each metronome separately, keeping the 
microphone stationary.
The microphone should be located at the origin of coordinates and the two
metronomes should be at angular location $0^\circ$.
The file prefixes and bpms in the text entry boxes should be chosen
to match the physical settings of the metronome.

\i After recording the pulse data for each metronome, you can play it 
back, calculate the corresponding pulse profile and period, calculate
the residuals, and detrend the residuals, by simply pressing the relevant
GUI buttons.
\een

This analysis produces two pulse profile data files 
(e.g., {\tt m208a\_profile.txt} and {\tt m184b\_profile.txt}) and the 
associated pulse periods for the two metronomes, which are needed for 
the double-metronome analysis described in the next section.

%%%%%%%%%%%%%%%%%%%%%%%%%%%%%%%%%%%%%%%%%%%%%%%%%%%
\section{Double-metronome analysis}
\label{s:analysis2}

The purpose of the double-metronome analysis is to experimentally
verify the expected $\rho_{12} \simeq \cos\zeta$ correlation 
coefficient for the timing residuals for the two metronomes when
the microphone is undergoing uniform circular motion.
This is analogous to real pulsar timing analyses looking for 
evidence of the Hellings and Downs correlation curve when correlating
timing residuals for pairs of Earth-pulsar baselines.

A screenshot of the GUI for this analysis, {\tt PTAdemo2GUI.py}, 
is shown in Figure~\ref{f:PTAdemo2GUI}.
The GUI has space for plots of:
(i) pulses from the two metronomes running simultaneously,
(ii) reference pulse profiles for the two metronomes, which were 
calculated using {\tt PTAdemo1GUI.py}, and
(iii) timing residuals for the two metronomes.
The GUI also has several text entry fields and buttons, whose functions
are described below:
\begin{figure}[hbtp!]
\begin{center}
\includegraphics[clip=true, angle=0, width=\textwidth]{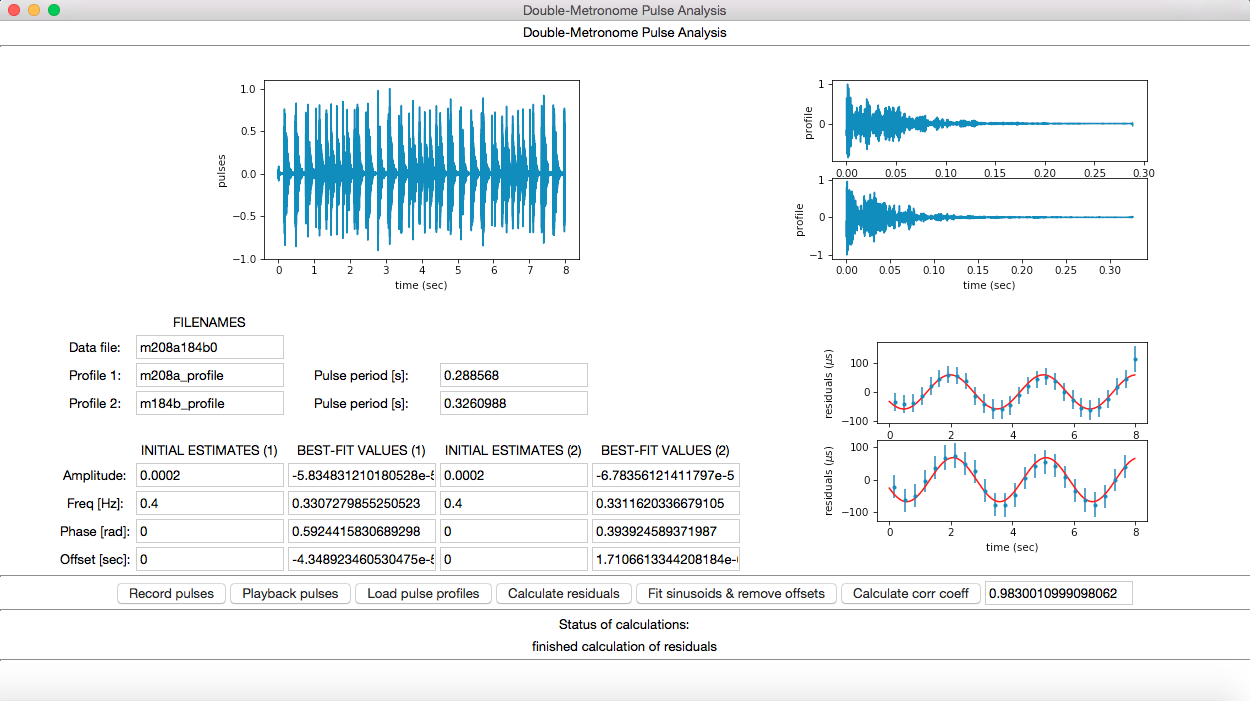}
\caption{Python GUI for the double-metronome analysis.
Plots, text entry fields, and buttons are described in the main text.}
\label{f:PTAdemo2GUI}
\end{center}
\end{figure}

\begin{description}

\i {\bf Record pulses}:
Records audio pulse data from the two metronomes running simultaneously, 
and then saves the corresponding timeseries to an ascii {\tt .txt} file 
with file prefix specified by the {\tt Data file} text entry box under 
the {\tt FILENAMES} label (default {\tt m208a184b0}).

\i {\bf Playback pulses}:
Plays back and plots the audio pulse data saved in
{\tt m208a184b0.txt}.

\i {\bf Load pulse profiles}:
Reads-in and plots the reference pulse profiles for the two 
metronomes, which were calculated by {\tt PTAdemo1GUI.py} and
were saved in ascii {\tt .txt} files specified by 
the {\tt Profile 1,2} text entry boxes under the {\tt FILENAMES} label
(default {\tt m208a\_profile} and {\tt m184b\_profile}).
The text entry boxes for the pulse periods should be filled with 
in with the values calculated by {\tt PTAdemo1GUI.py}.

\i {\bf Calculate residuals}:
Calculates the timing residuals as described previously
for {\tt PTAdemo1GUI.py}.

\i {\bf Fit sinusoids \& remove offsets}:
Simultaneously removes constant offsets and calculates best-fit 
sinusoids to the timing residuals for the two metronomes, using 
initial estimates for the amplitude, frequency, and phase of
the sinusoids, and the constant offset given in the text entry boxes labeled 
{\tt INITIAL ESTIMATES (1,2)} (defaults $2\times 10^{-4}$, $0.4~{\rm Hz}$,
0~radians, and 0~sec, respectively).
The constant offset arises from the arbitrariness of setting the 
timing residual of the pulse with the highest correlation to zero.
The best-fit parameter values calculated by {\tt Fit sinusoids \& remove offsets} 
are written to the text entry boxes labeled {\tt BEST-FIT VALUES (1,2)}.

\i {\bf Calculate corr coeff}:
Calculates the time-averaged correlation coefficient between 
the best-fit sinusoids for the two sets of timing residuals,
as described in Sec.~\ref{s:corrcoeff}.
Theoretically, the value of the correlation coefficient 
should equally $\cos\zeta$, where $\zeta$ is the separation angle 
between the line-of-sights to the two metronomes, 
as described in Sec~\ref{s:expectedcorrelation}.

\end{description}

\subsection{Steps for doing the analysis}

\ben

\i Start by placing both metronomes at the same angular location $0^\circ$
and at the same distance $L\sim 1~{\rm m}$ from the origin.
With both metronomes running simultaneously, record the audio data
while moving the microphone in uniform circular motion 
about the origin:
Typically, it is best to have 
$A\approx 10~{\rm cm}$ (=0.1~m) and period of oscillation 
$T_0\equiv 1/f_0\approx 2~{\rm s}$.
This leads to microphone-induced timing residuals of order 
$A/c_{\rm s}\approx 3\times 10^{-4}~{\rm s}$, where 
$c_{\rm s}=340~{\rm m/s}$ is the
speed of sound in air.
This timing precision turns out to be more than an order-of-magnitude larger than
the precision to which we can estimate the TOAs
of the metronome pulses, meaning that we can easily observe the
effect of microphone motion in the timing-residual data.

\i After recording the double-metronome data, you can play it back, 
load the pulse profiles, and calculate the residuals 
for each metronome.
The timing residuals induced by the microphone motion should be
sinusoidal and have large signal-to-noise ratio.
You should then fit sinusoids to the residuals for each metronome, 
adjusting the {\tt INITIAL GUESS} amplitudes, frequencies, and phases if 
necessary.  
(The initial guesses just have to be close, not exact.)
Finally, you should calculate the correlation coefficient, 
which should have a value very close to 1 for this case, since 
the two microphones are at the same angular location.

\i Repeat the above two steps but with metronome 2 at different
angular locations ($45^\circ$, $90^\circ$, $135^\circ$, $180^\circ$)
with respect to metronome 1 (which should always remain at $0^\circ$).
The motion of the microphone should be as similar as possible to that
for Step 1.
Change the name of the file prefix in the 
{\tt Data file} text entry box to {\tt m208a184bXX}, 
where {\tt XX} is {\tt 45}, {\tt 90}, {\tt 135}, {\tt 180}, 
to reflect the change in the angular
location of metronome 2.
You should find that the correlation coefficient is approximately
equal to $\cos\zeta$, where $\zeta=45^\circ$, $90^\circ$, $135^\circ$, $180^\circ$
is the angular separation of the two metronomes.

\een

This analysis produces data files ({\tt m208a184bXX.txt}, 
where {\tt XX} is {\tt 0}, {\tt 45}, {\tt 90}, {\tt 135}, {\tt 180}),
containing the double-metronome pulse timeseries.

%%%%%%%%%%%%%%%%%%%%%%%%%%%%%%%%%%%%%%%%%%%%%%%%%%%%%%%
\section{Discussion}
\label{s:discussion}

We have described a demonstration using two metronomes and a 
microphone that serves as an 
acoustical analogue of a Galactic-scale 
gravitational-wave detector, i.e.,
a pulsar timing array.
This demonstration also serves as an educational tool,
illustrating several techniques used in real
pulsar timing analyses, but in the simplified context of 
metronome pulses recorded by a microphone.
From our experience, we have found that the demonstration
is best suited for undergraduates or senior-level
high-school students who already have some familiarity
with basic physics and astronomy.
For less mathematically-inclined audiences, the 
mathematical discussion of the underlying data analysis
techniques needs to be reduced accordingly.
But the main idea that a common disturbance (in this case,
the microphone motion) can induce correlations in the 
pulse arrival times, and a graphical display showing 
the timing residuals from the two metronomes being 
shifted by an amount equal to their angular separation is 
accessible to nearly all audiences.

\subsection{Some caveats}
\label{s:caveats}

The tricky technical aspect of the double-metronome
analysis is to properly extract the pulse 
arrival times when the two metronomes are running 
simultaneously, producing pulses that can 
significantly overlap with one another.
The fact that the pulse profiles $p_I(t)$ for the two 
metronomes ($I=1,2$) 
are different for different tempo modes $a$ and $b$ 
is crucial for distinguishing the pulses from the 
two metronomes.
Still, the correlation functions 
$C_I(\tau)$ have several local maxima, and we 
need to find the largest local maxima in the 
vicinity of the expected pulse arrival times to 
determine the measured TOAs $\tau_{I}^{\rm meas}[i]$.
If the search window is not properly centered on the 
expected arrival time or if it includes a local maximum 
of the correlation function that doesn't correspond 
to the true arrival time of the pulse, then the 
returned measured TOA will deviate from its true 
value, thus causing errors in the corresponding timing 
residual and the subsequent fit to the residuals.
To help mitigate such problems, the routine that
calculates the measured TOAs currently uses 
an adaptive width for the search window, which 
increases in size if it originally does not
include a peak in the correlation function
(this is usually a sign that the window was not 
large enough to include the true pulse arrival time).

Even with this adaptive-search-window technique,
we sometimes do not get good agreement between the 
measured and theoretical
correlation coefficients for intermediate separation 
angles between the two metronomes, 
i.e., $\zeta$ close to 90~degrees.
A possible alternative 
reason for this might be reflections of the 
sound waves off of the table top or parts of the 
laptop, when using the laptop's internal microphone 
to do the recordings.
Recovery of the expected correlation is usually better 
if we use a USB microphone, which does not have many 
intervening parts to interfere with the sound waves.

\subsection{Possible use as an instructional laboratory
investigation}
\label{s:lab}

Although we have not tried to use this demonstration in its full form
as an instructional laboratory, we suspect that 
some variant of this demonstration might be useful 
for an undergraduate physics or astronomy lab.
We have used the single-metronome demonstration at public
outreach events and for a high-school Advanced Placement Physics 
demonstration with good success in getting students to
understand the fundamentals of pulsar timing based on
questions asked throughout. In a lab for more advanced
students, the usefulness of the full demonstration
comes in the form of learning the data analysis techniques of 
folding, matched-filtering, cross-correlation, etc., 
which are very general and have widespread
applicability in many branches of science.
Students who are comfortable with computer 
programming could be asked to code up their own
data folding and matched-filtering routines, etc., 
and apply them to the metronome pulse data.
Or the students could take the routines that 
already exist, but create their own customized
GUIs to perform 
custom analyses on other recorded sound files.
Of course, one could simply try to use the 
existing demonstration (more or less as is) 
as a lab, but we suspect that it would be 
best if it
were done as a ``communal effort", at least 
as far as 
the metronome data taking is concerned.
In other words, the two metronomes and single 
microphone would be shared amongst all lab 
groups, but each group would be responsible for 
performing one of the single-metronome 
or double-metronome analyses (e.g., a double-metronome
analysis for a specific angular separation).
Otherwise, there would be too much noise 
contamination from $\sim\!10$ pairs of 
metronomes running simultaneously!

\subsection{Enhancements under development}
\label{s:enhancements}

To make it easier for people who are not computer savvy 
to perform the demonstration, we are currently 
developing a  web-based interface for 
running the data analysis part of the demonstration.
This will eliminate the need for the demonstrator 
to have a working installation of all the requisite 
Python routines 
and Python packages on his/her own computer, 
and should simplify the operation of the GUIs.
Although for this scheme the data analysis will be 
done remotely
on the web server, the data taking will still be 
done locally using the two (physical) metronomes 
and e.g., a smartphone for recording the pulses.  
The sound files recorded by the smartphone would 
then need to be uploaded to the web server for the 
subsequent single-metronome and double-metronome
analyses.

Moving further in this direction, we also have 
an implementation of the metronome-microphone 
demonstration that exclusively uses readily-available
smartphones to drive the demonstration\citep{kv2013,mg2014,Osorio+2018}.
We have written a smartphone app, 
called {\em TableTopPTA}, which allows a 
smartphone to operate as either a metronome
or a microphone, as well as to perform all of the
data analysis calculations needed for the
single-metronome and double-metronome analyses.
The demonstration can then be done using just 
three smartphones, two of which run in 
metronome mode; the other running in microphone 
mode and performing the subsequent
data analysis calculations.
The app is written in Javascript and runs on Android 
smartphones; 
we have not yet written an iPhone version of 
the app.
The current code is available for public download from URL
\url{https://github.com/marcnormandin/tabletop_pta}.
Although some of the analysis routines for the 
smartphone app are not currently as up-to-date as 
those for the Python implementation of the 
demonstration, we have decided to make the code publicly 
available in case people want to experiment with 
what we currently have and possibly improve things in the process.

%%%%%%%%%%%%%%%%%%%%%%%%%%%%%%%%%%%%%%%%%%%%%%%%%%%%%%%
\begin{acknowledgments}
We would like to acknowledge support from NSF Physics Frontier Center
award number 1430284.
JDR and MN would also like to acknowledge support from NSF grant
PHY-1505861.

\end{acknowledgments}
%%%%%%%%%%%%%%%%%%%%%%%%%%%%%%%%%%%%%%%%%%%%%%%%%%%%%%%
%\appendix


\begin{thebibliography}{5}

\bibitem[Detweiler(1979)]{Detweiler1979} Detweiler, S.\ 1979, ``Pulsar Timing Measurement and the Search for Gravitational Waves'', \apj, 234, 1100 

\bibitem[Hellings \& Downs(1983)]{hd1983} Hellings, R.~W., \& Downs, G.~S.\ 1983, ``Upper Limits on the Isotropic Gravitational Radiation Background from Pulsar Timing Analysis'', \apjl, 265, L39 

%\bibitem[Riles(2013)]{Riles2013} Riles, K.\ 2013, ``Gravitational Waves: Sources, Detectors and Searches'', Progress in Particle and Nuclear Physics, 68, 1 

%\bibitem[Danzmann \& LISA study Team(1996)]{Danzmann1996} Danzmann, K., \& LISA Study Team, 1996, ``LISA: Laser Interferometer Space Antenna for Gravitational Wave Measurements'', Classical and Quantum Gravity, 13, A247 

\bibitem[Bizouard et al.(2013)]{PTACQG} Bizouard, M.~A., Jenet, F., Price, R., \& Will, C.~M.\ 2013, ``Pulsar Timing Arrays'', Classical and Quantum Gravity, 30, 220301 

%\bibitem[Guzzetti et al.(2016)]{Guzzetti+2016} Guzzetti, M.~C., Bartolo, N., Liguori, M., \& Matarrese, S., 2016, ``Gravitational Waves from Inflation'', Riv. Nuovo Cim. 39, 9, 399-495

\bibitem[Lorimer \& Kramer(2012)]{handbook} Lorimer, D.~R., \& Kramer, M.\ 2012, {\it Handbook of Pulsar Astronomy}, by D.~R.~Lorimer , M.~Kramer, Cambridge, UK: Cambridge University Press, 2012

\bibitem[Hobbs et al.(2012)]{Hobbs+2012} Hobbs, G., Coles, W., Manchester, R.~N., et al.\ 2012, ``Development of a Pulsar-based Time-scale'', \mnras, 427, 2780 

\bibitem[Cordes \& Downs(1985)]{cd1985} Cordes, J.~M., \& Downs, G.~S.\ 1985, ``JPL Pulsar Timing Observations. III. Pulsar Rotation Fluctuations'', \apjs, 59, 343 

\bibitem[Lam et al.(2017)]{NG9EN} Lam, M.~T., Cordes, J.~M., Chatterjee, S., et al.\ 2017, ``The NANOGrav Nine-Year Data: Excess Noise in Millisecond Pulsar Arrival Times'', \apj, 834, 35

\bibitem[Lentati et al.(2016)]{IPTADR1noise} Lentati, L., Shannon, R.~M., Coles, W.~A., et al.\ 2016, ``From Spin Noise to Systematics: Stochastic Processes in the First International Pulsar Timing Array Data Release'', \mnras, 458, 2161 

\bibitem[Cordes(2013)]{CordesCQG} Cordes, J.~M.\ 2013, ``Limits to PTA Sensitivity: Spin Stability and Arrival Time Precision of Millisecond Pulsars'', Classical and Quantum Gravity, 30, 224002

\bibitem[Condon \& Ransom(2016)]{ERA} Condon, J.~J., \& Ransom, S.~M.\ 2016, ``Essential Radio Astronomy'', Princeton University Press

\bibitem[Estabrook \& Wahlquist(1975)]{ew1975} Estabrook, F.~B., \& Wahlquist, H.~D.\ 1975, ``Response of Doppler Spacecraft Tracking to Gravitational Radiation'', General Relativity and Gravitation, 6, 439 

\bibitem[Abbott et al.(2016)]{GW150914} Abbott, B.~P., Abbott, R., Abbott, T.~D., et al.\ 2016, ``Observation of Gravitational Waves from a Binary Black Hole Merger'', Physical Review Letters, 116, 061102 

\bibitem{LIGOPapers} \url{https://www.ligo.caltech.edu/page/detection-companion-papers}

\bibitem[Abbott et al.(2017)]{GW170817} Abbott, B.~P., Abbott, R., Abbott, T.~D., et al.\ 2017, ``Observation of Gravitational Waves from a Binary Neutron Star Inspiral'', Physical Review Letters, 119, 161101 

\bibitem[Jenet \& Romano(2015)]{jr2015} Jenet, F.~A., \& Romano, J.~D.\ 2015, ``Understanding the Gravitational-Wave Hellings and Downs Curve for Pulsar Timing Arrays in Terms of Sound and Electromagnetic Waves'', \ajp, 83, 635

%\bibitem[Dethlefsen(1961)]{Dethlefsen1961} Dethlefsen, E.~S.\ 1961, ```Gravity' Demonstration Using a Magnetic Field'', \ajp, 29, 549

\bibitem[Rubbo et al.(2007)]{Rubbo+2007} Rubbo, L.~J., Larson, S.~L., Larson, M.~B., Ingram, D.~R.\ 2007, ``Hands-on Gravitational Wave Astronomy: Extracting Astrophysical Information from Simulated Signals'', \ajp, 75, 597

\bibitem[Newburgh(2008)]{Newburgh2008} Newburgh, R.\ 2008, ``A Demonstration of Einstein's Equivalence of Gravity and Acceleration'', European Journal of Physics, 29, 2

\bibitem[Ford et al.(2015)]{fsa2015} Ford, J., Stang, J., \& Anderson, C.\ 2015, ``Simulating Gravity: Dark Matter and Gravitational Lensing in the Classroom'', The Physics Teacher, 53, 557

\bibitem[Burko(2017)]{Burko2017} Burko, L.~M.\ 2017, ``Gravitational Wave Detection in the Introductory Lab'', The Physics Teacher, 55, 288


\bibitem[Farr et al.(2012)]{Farr+2012} Farr, B., Schelbert, G., \& Trouille, L.\ 2012, ``Gravitational Wave Science in the High School Classroom'', \ajp, 80, 898

\bibitem[Kassner(2015)]{Kassner2015} Kassner, K.\ 2015, ''Classroom Reconstruction of the Schwarzschild Metric'', European Journal of Physics, 36, 6

\bibitem[Mathur et al.(2017)]{Mathur+2017} Mathur, H., Brown, K., \& Lowenstein, A.\ 2017, ``An Analysis of the LIGO Discovery Based on Introductory Physics'', \ajp, 85, 676


\bibitem[Kaur et al.(2017)]{Kaur+2017} Kaur, T., Blair, D., Moschilla, J., Stannard, W., \& Zadnik, M.\ 2017, ``Teaching Einsteinian Physics at Schools: Part 1, Models and Analogies for Relativity'', Physics Education, 52, 6

\bibitem[Hilborn(2018)]{Hilborn2018} Hilborn, R.~C.\ 2018, ``Gravitational Waves from Orbiting Binaries Without General Relativity'', \ajp, 86, 186

%FOO 

\bibitem[Cordes \& Shannon(2010)]{cs2010} Cordes, J.~M., \& Shannon, R.~M.\ 2010, ``A Measurement Model for Precision Pulsar Timing'', arXiv:1010.3785

\bibitem[Taylor(1992)]{Taylor1992} Taylor, J.~H.\ 1992, ``Pulsar Timing and Relativistic Gravity'', Royal Soc. of London Phil. Trans. Series A, 341, 117 

\bibitem[Lam et al.(2016)]{NG9WN} Lam, M.~T., Cordes, J.~M., Chatterjee, S., et al.\ 2016, ``The NANOGrav Nine-Year Data Set: Noise Budget for Pulsar Arrival Times on Intraday Timescales'', \apj, 819, 155

\bibitem{FourierFootnote} The Fourier transform $\tilde y(f)$ of the
time-series $y(t)$ is defined by
$\tilde y(f) = \int dt\> y(t)e^{-i2\pi f t}$ 
or, equivalently,
$y(t) = \int df\> \tilde y(f)e^{i2\pi f t}$.
%\bibitem{footnotes}%It is necessary to process a file twice to
%get the counters correct. \ajp\ does not use footnotes.

\bibitem[Kuhn \& Vogt(2013)]{kv2013} Kuhn, J., Vogt, P.\ 2013, ``Applications and Examples of Experiments with Mobile Phones and Smartphones in Physics Lessons'', Frontiers in Sensors, 1, 4


\bibitem[Martinez \& Garaizar(2014)]{mg2014} Martinez, L. \& Garaizar, P.\ 2014, ``Learning Physics Down a Slide: A Set of Experiments to Measure Reality Through Smartphone Sensors'', International Journal of Interactive Mobile Technologies, 8, 3

\bibitem[Osorio et al.(2018)]{Osorio+2018} Osorio, M., Pereyra, C.~J., Gau, D.~L., \& Laguarda, A.\ 2018, ``Measuring and characterizing beat phenomena with a smartphone'', European Journal of Physics, 39, 2


%\bibitem{}

\end{thebibliography}
\end{document}